# Active Galactic Nuclei and their role in Galaxy Formation and Evolution

**13 February 2009**

Steve Kraemer (CUA), Rogier Windhorst (ASU), Kenneth G. Carpenter (NASA-GSFC), Mike Crenshaw (GSU), Martin Elvis (CfA), and Margarita Karovska (CfA)

For more Information, please contact: Professor Steve Kraemer
Dept. of Physics
200 Hannan Hall
Catholic University of America
Washington, DC 20064
kraemer@cua.edu
301-286-8301

*Scientific Background:*

Active Galactic Nuclei (AGN) have been studied over the entire electromagnetic spectrum, using both ground-based and satellite observatories. The basic aspects of the phenomenon are generally agreed upon. As shown in Figure 1, the AGN can influence its host galaxy at scales ranging up to Mpcs. However, in order to fully probe the critical role of the AGN in galaxy formation/evolution, new capabilities are required: specifically, sub-milliarcsecond (sub-mas) optical/UV imaging that can only be achieved with space-based, long-baseline (0.5-1.0 km) observatories, such as the generic Ultraviolet Optical Interferometer (UVOI) under consideration for the decade of the 2020's and beyond (see 2006 NASA Astrophysics Strategic Plan).

A fraction of galaxies harbor powerful non-stellar energy sources, AGN, at their gravitational centers. AGN emit radiation at all energies and span a huge range in luminosity, from Low Luminosity AGN and LINERs, to Seyfert galaxies, and, finally, QSOs. The source of the AGN's power is believed to be accretion of matter by a supermassive black hole (SMBH) in the center of the host galaxy's nuclear bulge. Matter in the host galaxy, having lost angular momentum, spirals towards the SMBH, forming an accretion disk.

Viscous forces heat the disk to ~ several x $10^4$ K. A hot (>$10^6$ K) coronal forms above the disk, most likely heated via magnetic reconnection, Thermal photons from the disk are up-scattered by relativistic electrons in the corona, producing EUV–X-ray continuum radiation. Also, relativistic particles are accelerated and collimated along magnetic fields in the inner parts of the disk and are ejected along the rotation axis of the SMBH/disk system, forming the extended jets observed in some AGN.

The optical and UV spectra of AGN are characterized by broad emission lines. Doppler-broadened permitted lines, with full width at half maximum (FWHM) > several 1000 km $s^{-1}$ are thought to form in dense gas in a region within tens of light days from the central black hole, referred to as the "Broad Line Region (BLR)", while forbidden lines, with FWHM< 1000 km $s^{-1}$, form in lower density gas in the "Narrow-Line Region (NLR)", which may extend from 1 pc to several kpcs (depending on luminosity, see Bennert et al. 2002). Furthermore, AGN are divided into Type 1s, which show broad permitted lines and non-stellar optical continua and Type 2s, which have permitted and forbidden lines of similar widths and continua dominated by the host galaxy (Khachikian & Weedman 1974). The discovery of polarized broad lines in the spectra of Type 2 Seyfert galaxies (Miller & Antonucci 1983) led to the unified model which posits that the two types are intrinsically the same but that our line-of-sight to the BLR and accretion disk in Type 2s is blocked by a dusty circumnuclear torus (Antonucci 1993).

**There are several key questions as to the nature and origin of AGN that can be addressed only by probing their central regions with sub-mas angular resolution at UV/optical wavelengths. These include 1) what initiates the active phase, 2) the duration of the active phase, and 3) the effect of the AGN on the host galaxy.** Notably, from careful studies of the rotation curves of nearby galaxies, it is believed that all galaxies with massive bulges possess the "engine" of the AGN, i.e. a SMBH in their centers (Kormendy & Richstone 1995). Remarkably, the SMBH mass is roughly proportional to the galaxy bulge mass over more than 4 orders of magnitude (Ferrarese & Ford 2005): $M_{SMBH} \approx 0.002$ (±0.4 dex) x $M_{bulge}$. For example, in a giant Elliptical galaxy in the present epoch, the SMBH may have accumulated up to $10^{10}$ $M_{sun}$ over a Hubble time.

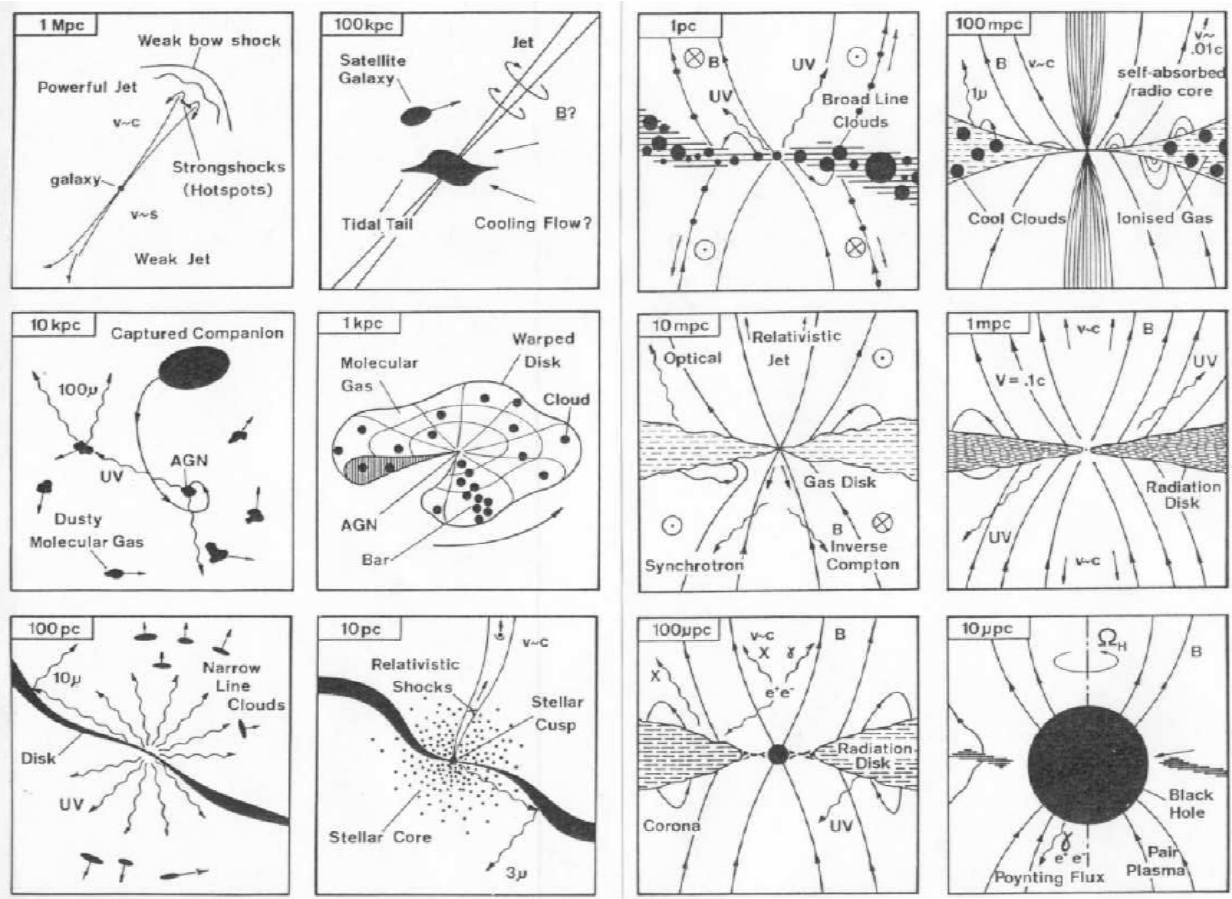

**Fig.1**: Summary of how AGN affect their surroundings over 12 orders of magnitude in size: from relativistic radio jets at Mpc scales (upper left) to the supermassive black-hole (SMBH) and its surrounding accretion disk at AU or micro-pc scales (lower right). Starting from the upper left, each next panel is expanded by a factor of 10. The SMBH is well visible in the lower right two panels, and the inner accretion disk and torus in the right 6 panels (pc-AU scales). The outer AGN accretion disk and the escaping relativistic jet are well visible in the left 6 panels (Mpc-pc scales), with the galaxy itself shown in the 100-kpc panel (2nd from upper left). Figure from R. Blandford in Active Galactic Nuclei (1990; Springer Verlag, Berlin).

This relationship suggests that the growth of the SMBH has kept pace with the process of hierarchical galaxy assembly (e.g., Cohen et al. 2006). The trigger for the build-up of the bulge/SMBH is thought to be major galaxy mergers, which, in turn feed the central accretion disk and initiate a burst of star formation. **The current paradigm posits that the accumulation of matter in the bulge is halted by the effect of the AGN, i.e. "AGN feedback'" (e.g. Kauffmann & Haehnelt 2000).** One can divide the AGN feedback into 1) radio-mode feedback, in which the onset of the jet disrupts gas cooling in the galaxy's halo (Croton et al. 2006), and 2) quasar-mode feedback, in which the powerful radiation emitted by the AGN drives material out of the galactic bulge (Hopkins et al. 2006). There appears to be a significant (1-2 Gyr) time delay between each merger and the ignition of the AGN (Hopkins et al. 2006), in part due to the time it takes material in the disk to reach the vicinity of the SMBH. Statistical studies suggest that the lifetimes of AGN are ~ several x$10^7$ yrs (e.g. Kauffmann & Haehnelt 2000), or roughly 1% of the typical major merger timescales. Hence, the bulge/SMBH is built-up via a number of major mergers, between which the AGN is in a quiescent state (during which there may be some low-level of activity).

Approximately 50% of Type 1 AGN show blue-shifted absorption lines in their UV spectra (Crenshaw et al.1999; Ganguly & Brotherton 2008), indicative of mass outflow. In order to power the AGN, mass must be driven from the vicinity of the SMBH to remove angular momentum from the accretion disk (e.g., Blandford & Payne 1982). Hence, mass outflow is an essential element in the energetics of AGN (Elvis 2000). In Seyfert galaxies, the outflow appears to be driven by magneto-hydrodynamic processes (e.g., Kraemer et al. 2005; Turner et al. 2005), while the more energetic outflows detected in QSOs are more likely radiatively driven (e.g., Arav et al. 1995). For Seyferts, and other low-luminosity AGN, the mass loss rates and kinetic luminosities are much lower than required for AGN feedback. However, QSOs show much more energetic, and optically thick, outflows (e.g. BAL QSOs). In fact, outflow velocities of several x $10^4$ km s$^{-1}$ have been detected (e.g. Ganguly & Brotherton 2008). Hence, the QSO outflows share the characteristics of the initial AGN turn-on that is thought to have occurred during the bulge formation.

For the most part, what has been learned about mass outflow has come from absorption-line studies. From these, the column densities, radial velocities, and, in some cases, radial distances for the absorbers have been determined (e.g. Arav et al. 2008). However, in order to determine mass-loss rates and kinetic luminosities, one must determine the total amount of material in the outflow, specifically the global covering factors of the absorbers. One way to determine the covering factors is by spectral imaging of the outflowing gas. Using HST/STIS long slit spectra, Crenshaw & Kraemer (2007) were able to identify in emission one component of the outflow in the Seyfert 1 galaxy NGC 4151, but only were able to resolve structure of ~10 pc in extent, while the bulk of the outflow is ~ 0.1 pc from the AGN. However, with 0.1 mas resolution achievable via a UVOI (e.g., Carpenter et al. 2008), we can resolve the inner 10 light days in local AGN such as NGC 4151(see Figure 2). This is just outside the BLR region (Clavel et al. 1990), and, since the absorbers appear to cover the BLR emission (Crenshaw et al. 1999), this allows us to probe the region in which the bulk of the outflowing mass exists.

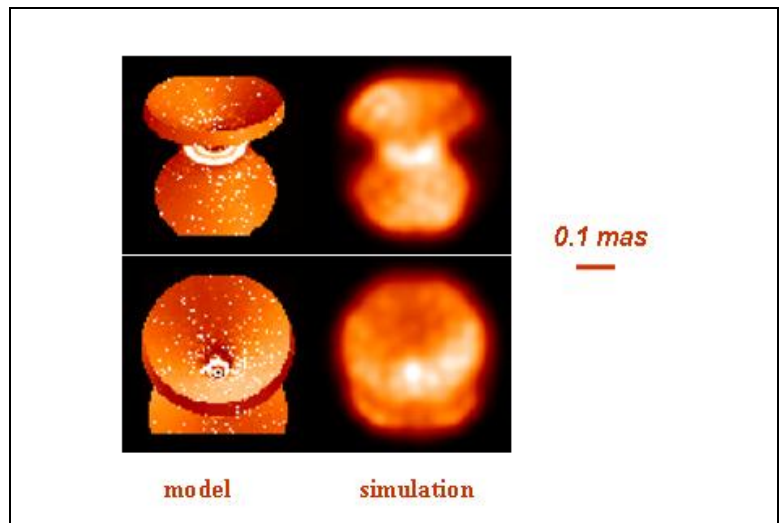

Fig.2 : Elvis (2002) Model (left) and simulation (right) of an UVOI (500 m baseline) observation of the inner ~ 100 light days of nearby AGN in CIV 1550 Å. The simulation demonstrates that we will be able to resolve the structure of the circumnuclear gas which will allow us to determine the origin and physical characteristics of mass outflow.

More importantly, scaling up to redshifts of $z$ ~ 0.5, we will be able to resolve structure of ~ 1 pc. Hence, we will be able to map the global extent of the outflows in intermediate redshift QSOs, thereby obtaining accurate constraints on their energetics.

*Based on the strong UV resonance lines that are the main signatures of mass outflow, the best way to map their global extent is via Ly-α and C IVλ1550.* Although these lines dominate the BLR spectra of AGN, with UVOI resolution we will be able to remove the unresolved BLR and directly detect the Ly- α and C IV from the surrounding gas.

*Scientific Goals*:
To summarize, the main science goals that can be achieved with UVOI interferometry are as follows:

**1. Constraining the dynamics of AGN feedback.** As noted above, UVOI spectral imaging will be used to map the outflows in intermediate redshift QSOs. This will provide new insight into AGN feedback with which models of hierarchical build-up of galactic bulges can be tested.

**2. Probing the BLR/NLR transition regions in AGN**. Based on high-resolution UV spectroscopy, the bulk of the absorbing material in AGN lies outside the BLR, most likely in the BLR/NLR transition region. Resolving this zone will allow us to trace the launch point of the mass outflow. This will give us new insight into how gas is ejected from AGN.

**3. New insights regarding the structure of the torus**. One possible source for the outflow is ablation of material from the inner face of the torus. Note, that sub-mas resolution will allow us to probe the sublimation regions over a fairly large range in redshift, which is relevant to both the torus structure and the BLR/NLR transition region.

**4. Jet formation/collimation**. Although we will not have the ability to trace the jet back to the accretion disk, this is also the region where the relativistic jet is collimated for its journey of over 7 orders of magnitude in distance into the intergalactic medium. Furthermore, the UV/optical spectral images will be complementary to the ground-based VLBI and IR observations.

**5. Precise measurement of the average opening angle of AGN at sub-pc scales**. A UVOI will provide a direct measure of the average cone opening angle of AGN, and if this number is close to $45^{\circ}$ this would be a strong direct confirmation of the AGN unification picture. If significantly different from $45^{\circ}$, then the AGN unification picture may well be incorrect, posing a severe challenge to the field.

*Future Directions:*
An additional science goal of a UVOI will be standard-length distance measurements. Mapping the 3-D geometry of the Universe involves measurement of the large "cosmic" scale distances of high redshift sources such as distant supernovae and quasars. Standard distance measurements use relative distance estimation, e.g., the brightness of supernovae of type SNIa at $z$ ~1.5 as "standard candles" (Perlmutter et al. 1999). Recently Elvis & Karovska (2002) proposed an absolute method for estimating distances to quasars at different redshifts using long-baseline interferometry of quasar BLRs, which would provide an independent check on the standard technique. This geometric method uses the size of the quasar BLR from reverberation mapping (Peterson et al. 2004) combined with interferometric measurements of the angular diameter of the emitting region to derive the distance to the quasar. The quasar broad emission lines (v~5000 – 10,000 km s$^{-1}$) originating in the BLR gas clouds respond to changes in the continuum source in the center by changing their intensity (~20% in the UV) with a time-lag of a few days to years. This time-lag is induced by the light travel time from the continuum source. For low redshift quasars the size of the BLRs is ~10 light days corresponding to an angular size of a fraction of a mas (Figure 3). When compared to relative distance estimators this method is much less dependent on physical models and on changes in the fundamental constants (other than c, the speed of light) because it uses a standard-length measurement approach rather than a standard-candle approximation. Although there are currently a small number of AGN with BLRs that can be resolved with 0.1 mas imaging, the first UVOI's will be pathfinders for future space missions that will extend another factor 10-10³ in spatial resolution, which will not only resolve the BLR, but will allow us to reach the Schwarzschild radius of $10^8$-$10^9$ $M_{sun}$ SMBH's. These will conceivably yield the first direct measures of black-hole shadows.

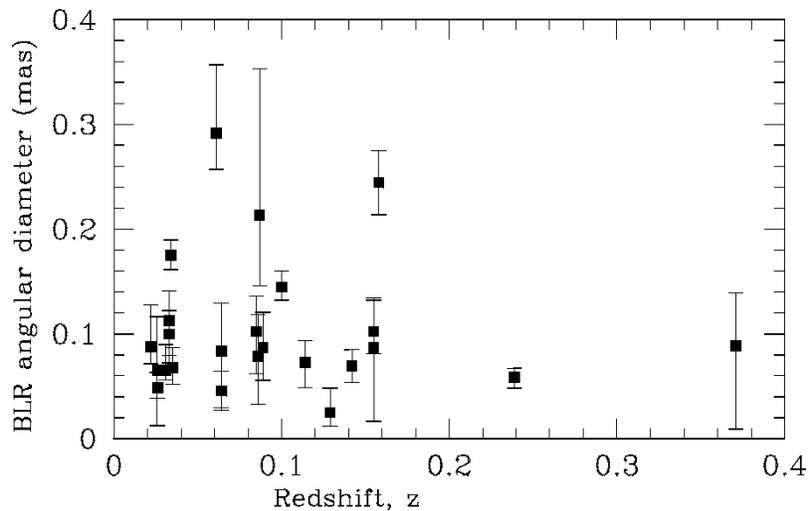

**Figure 3**: Angular diameters for the H$\alpha$ and H$\beta$ BELRs of nearby active galaxies, assuming $H_0$=65 km s$^{-1}$Mpc$^{-1}$. (Peterson et al. 2004, Kaspi et al. 2000)

*Summary*

We have presented the compelling new AGN science that can be accomplished with sub-mas resolution. In particular, such observations would enable us to constrain the energetics of the AGN "feedback" mechanism, which is critical for understanding the role of AGN in galaxy formation and evolution. These observations can only be obtained by long-baseline interferometers or sparse aperture telescopes in space, since the aperture diameters required are in excess of 500 m – a regime in which monolithic or segmented designs are not and will not be feasible and because these observations require the detection of faint emission near the bright unresolved continuum source, which is impossible from the ground, even with adaptive optics. Two mission concepts which could provide these invaluable observations are NASA's Stellar Imager (SI; Carpenter et al. 2008 & http://hires.gsfc.nasa.gov/si/) interferometer and ESA's Luciola (Labeyrie 2008) sparse aperture hypertelescope. For example, the small field-of-view of SI will permit us to observe close to the AGN while minimizing the scattered light from the central source. However, another possibility would be to include a nulling coronagraph instrument on either of these telescopes.